\documentclass{aa}
\usepackage{graphicx}
\usepackage{rotating}
\usepackage{afterpage}
\usepackage[sort&compress]{natbib}
\bibpunct{(}{)}{;}{a}{}{,}
\begin{document}

   \title{The optical morphology of A3667 re-examined}

   \author{M. Johnston-Hollitt
          \inst{1}
          \and
          R. W. Hunstead\inst{2}
          \and
          E. Corbett\inst{3}
          }

   \offprints{M. Johnston-Hollitt}

   \institute{School of Maths \& Physics, University of Tasmania, 
              Private Bag 21, Hobart, Tas 7001, Australia \\
              \email{Melanie.JohnstonHollitt@utas.edu.au}\\
         \and
             School of Physics, University of Sydney, NSW 2006, Australia\\
         \and 
             Anglo-Australian Observatory, P.O. Box 296, Epping NSW 1710, 
             Australia\\
             }

   \date{Received XXXX XX, 200X; accepted XXXXX XX, 200X}

   \abstract{The galaxy cluster A3667 was observed using the Two-degree Field (2dF)
multifibre spectroscopic system on the Anglo-Australian Telescope in a
program designed to examine the velocity structure in the
region. Specifically, we sought evidence from the optical data for the
putative cluster merger believed to be responsible for the observed
radio and X-ray emission. We present 184 new redshifts in the region,
of which 143 correspond to member galaxies of A3667. We find the
cluster velocity distribution to be well modelled by a single
Gaussian in agreement with previous results. 
In addition, new redshift-selected isodensity plots
significantly reduce the prominence of the previously reported
subgroup to the north-west of the main cluster. Instead, we find 
the galaxy distribution to be
elongated and well mixed, with a high velocity dispersion and no significant
evidence for substructure. These
results are consistent with the axis of the proposed merger being close to the 
plane of the sky.

   }

   \maketitle
%

\section{Introduction}
\label{intro}

The ACO galaxy cluster, A3667 \citep{ACO}, is a rich, X-ray-luminous
cluster, which is arguably one of the most interesting and,
consequently, most studied clusters in the southern sky.  It has the
distinction of being one of only a handful of galaxy clusters to have
a cold gas front at the centre \citep{Vikhlinin01} and the only
cluster yet seen with two bright, diffuse, radio emission regions
straddling the X-ray gas \citep{Rottgering97}.  Our physical picture
of the cluster is becoming clearer as more multiwavelength data are
obtained, but many basic questions about the dynamical history of
A3667 remain unanswered.

A3667 is relatively close, with a published mean redshift of 0.055 and
a large velocity dispersion ($\sigma_{\rm v}$) of 1000--1500 km s$^{-1}$
depending on the number of galaxies used and the redshift limits
applied \citep{Melnick81, Proust88, Sodre92, Girardi98}.  The Abell 
richness class is 2 and the
Bautz-Morgan type is I--II \citep{ACO}, indicating a high galaxy
density in this region. \citet{Proust88} presented evidence for a
bimodal spatial distribution of galaxies selected by apparent
magnitude, and a similar conclusion was reached by \citet{Sodre92}.

{\it ROSAT\/} observations revealed a high X-ray luminosity $L_X$,
consistent with the $L_X-\sigma$ relation \citep{Xue00}, of $8.74 \times 10^{44}\,
h^{2}$ erg s$^{-1}$ in the 0.4--2.4 keV range \citep{Ebeling96},
making it one of the brightest X-ray sources in the southern
sky\footnote{$H_0=100h$ km\,s$^{-1}$\,Mpc$^{-1}$}. Further, the X-ray
isophotes are distorted in the direction of the reported bimodal
optical distribution and observations from the {\it Einstein\/}
satellite are reported as showing clear evidence of substructure
\citep{Sodre92}. Observations of the temperature of the X-ray gas with
ASCA by \citet{Markevitch99} showed the central part of the cluster to
be cool and the X-ray surface brightness profile along the elongation
axis suggests a shocked region is present.  Recent Chandra
observations confirmed this interpretation, revealing the central part
of the cluster to contain a cold gas front moving through the warm
intracluster medium \citep{Vikhlinin01}.

Perhaps the most dramatic features of A3667 are the two extended,
symmetrically located regions of diffuse radio emission
\citep{Rottgering97}, first detected in 843 MHz images obtained with
the Molonglo Observatory Synthesis Telescope, and then
confirmed and studied in detail at 1.4 and 2.4 GHz with the Australia
Telescope Compact Array, ATCA \citep{mjh03}.  The combination of
these features has been interpreted as evidence that the cluster is
observed in a post-merger state after close core passage 
\citep{Rottgering97, KHB, Roettiger99, Markevitch99, Vikhlinin01}.

Given the strong indicator of merger activity present in the radio, 
X-ray and previous optical data we undertook a set of spectral
observation of the objects in the cluster to further test 
for signatures of a cluster-wide merger in
the optical population of A3667.

In Section 2 we discuss the signatures of mergers that have been
observed in galaxy clusters in the optical and present a detailed summary of
the previous optical data collected for A3667. Sections 3 and 4
discuss the Two-degree Field (2dF) observations and data reduction. The 
major results are discussed in Section 5 and the discussion and conclusions 
are presented in Section 6.

\section{Previous Optical Observations of A3667}
\label{prevoptical}

\citet{Melnick81} obtained redshifts for 16 member galaxies of
A3667. Following this, \citet{Proust88}, hereafter referred to as P88,
obtained a further 31 redshifts for galaxies in the field of A3667,
which they combined with the previous results.
Unfortunately, this combined sample has since been found to contain some unreliable redshift
values \citep{Sodre92}.  
P88 also posited a bimodal spatial distribution of galaxies in the
field of A3667 based on an isodensity plot of 423 galaxies with
$m_{\rm B} \leq 19$.
Two groupings can be seen in this plot.  
The main group was assigned a redshift of 0.0550, being that of the
galaxy closest to the centre of A3667, while the subgroup to the
north-west was suggested to be at $z=0.0563$, based on the redshift of
a nearby galaxy. It should be noted that nearly 90\% of the galaxies
in this analysis had unknown redshifts and there is an inevitable
contribution from foreground and background objects.

\citet{Sodre92} expanded on the P88 data by measuring redshifts for
128 galaxies in the vicinity of A3667.
This sample, hereafter referred to as S92, included reobservation of
all but five of the galaxies in the P88 sample and revealed some large
discrepancies in redshift. As neither study published spectra this has
made it difficult to determine which was the more reliable; in general
the S92 sample has smaller quoted errors and is assumed to be more
reliable. Other data in the literature found using a search with the
NASA Extragalactic database (NED) have provided a further 28
redshifts for objects in the S92 sample, bringing the total number of
redshifts available in the region of A3667 to 161. Of these, 155 fall
in the range $0.044 \leq z \leq 0.068$ and are likely to be cluster
members.

\citet{Girardi96} studied A3667 as part of a sample of 37 galaxy
clusters used to examine the relationship between velocity dispersion,
$\sigma_{\rm v}$, and X-ray temperature, $T$. They argued that since
both $\sigma_{\rm v}$ and $T$ were related to a cluster's
gravitational potential there would only be a strong correlation
between these quantities for clusters in which both gas and galaxies
were in dynamical equilibrium with the cluster potential.  Using the
123 redshifts from the S92 sample, \citet{Girardi96} examined A3667
for substructure using both a velocity distribution and velocity
gradient analysis. They found neither significant multiple peaks in the
velocity distribution, nor a velocity gradient. Using the method of
\citet{Dressler88} they computed the probability of substructure as
0.697 where a result above 0.990 was considered significant.  Despite
the lack of statistical evidence, their isodensity plot was
reminiscent of the P88 plot and suggested some substructure to the
northwest of the cluster centre. The velocity dispersion was
determined to be 1208$^{+95}_{-84}$ km s$^{-1}$.

Expanding on their earlier work, \citet{Girardi98} used 154 of the
then known 155 redshifts for A3667 to compute a maximum cluster radius
of $2.22 h^{-1}$ Mpc, a mean galactocentric redshift of 0.0566 and a
global velocity dispersion of $971^{+62}_{-47}$ km
s$^{-1}$. \citet{Girardi98} also obtained a virialised radius of $1.94
h^{-1}$ Mpc in which 152 of the galaxies with previously measured
redshifts are contained. The total virialized mass was calculated to
be $15.98^{+2.18}_{-1.71}\times 10^{14}$ M$_{\odot}$.  They further
concluded that the velocity distribution was isotropic.  In comparison
\citet{KHB} give the X-ray radius of the cluster to be $R_{X} = 0.80
h^{-1}$ Mpc and the X-ray mass at $M_{X} = 3.50^{+0.30}_{-0.40}\times
10^{14}$ M$_{\odot}$.  \citet{Girardi98} obtain an optical mass of
$M_{O} = 4.89^{+0.67}_{-0.52}\times 10^{14}$ M$_{\odot}$ within this
X-ray radius.  Assuming that the cluster galaxies follow the
gravitational potential, the good agreement between these mass
estimates strongly suggests that the cluster is in dynamical
equilibrium \citep{Girardi98}.

The mean redshifts and velocity dispersions obtained in the studies
cited above are summarized in Table \ref{tab:prevopt}.

\begin{table}
\caption{Mean redshifts and velocity dispersions for A3667.}

\begin{tabular}
{lcll}
\hline
\noalign{\smallskip}
Reference & N     & Mean $z$    &$\sigma_{\rm v}$ (km\,s$^{-1}$) \\
\noalign{\smallskip}
\hline
\\[-3mm]
\citet{Proust88} & 46$^{\mathrm{*}}$  & 0.054$\pm$0.003 & 1606$\pm$288\\
\citet{Sodre92}  & 120 & 0.0553$\pm$0.0004 & 1197$\pm$79\\
\citet{Girardi96}& 123 &        & 1208 $^{+95}_{-84}$\\
\citet{Girardi98}& 154 & 0.0566 & 971$^{+62}_{-47}$\\
This paper& 231 & 0.0555$\pm$0.0003\rlap{$^{\ddagger}$} & 1102$\pm$46\\[1mm]
\hline
\end{tabular}

\begin{list}{}{}
\item[$^{\mathrm{*}}$]
From the table in \citet{Proust88}.
\item[$^{\ddagger}$]
The mean of all 231 available redshifts; the best-fit Gaussian to
the data gives $\overline{z}=0.0553$ (see Section 6).
\end{list}
\label{tab:prevopt}
\end{table}

\section{2dF Observations}

In order to increase the redshift sample and examine the distribution
of galaxy types in the cluster, observations were undertaken with the
2dF spectrographic system \citep{Bailey02} on the AAT. SuperCOSMOS plate scans
were used to obtain a list of suitable galaxies with $15.5 \leq m_{\rm
B} \leq 18.3$. As there is a marked discrepancy between the redshift
values for some objects observed by P88 and S92 we decided to include
a subset of the S92 catalogue in the target sample to check for any
systematic differences in redshift. The optical counterparts of the
known radio galaxies in the region \citep{mjh03} were also
selected. This produced a final configuration of 338 target objects,
with a further 49 fibres assigned to guide stars and sky.

The observations were scheduled on 2001 July 20, using the 270R and
316R gratings. A total integration time of 1.5 hours was obtained in
three 30-minute integrations to allow cosmic ray removal via a
sigma-clipping algorithm. This gave a wavelength coverage of
approximately 3600--8000 \AA, varying slightly between the two
spectrographs due to the slightly different dispersions of the
gratings and from fibre to fibre on each spectrograph. The instrumental resolution was 9.9 \AA\ FWHM.

\section{Data Reduction}
\label{res6}

Data were reduced using the \verb+2dfdr+ pipeline software
package, which was developed for the 2dF Galaxy redshift survey
\citep{colless98, colless01}.  A full description of the
semi-automated extraction and reduction procedure is given in these
references.

The software performs two independent redshift estimations depending
on the emission and absorption properties of the spectra.  For the
absorption spectra we used the standard technique of cross-correlation
\citep{Tonry79} with a set of eight absorption templates, five
galaxies and three stars.  The galaxies used and their morphological
types, as listed in \citet{colless01}, are: (1) NGC 3379 (E), (2) NGC
4889 (cD), (3) NGC 5248 (Sbc), (4) NGC 2276 (Sc) and (5) NGC 4485
(Sm/Im). The stars used and their spectral classes are (6) HD 116608
(A1V), (7) HD 23524 (K0V) and (8) BD05$^\circ$1668 (M5V).

Spectra are prepared for cross-correlation via a six-step process that
includes: continuum-subtraction; clipping strong emission lines;
rebinning to a logarithmic scale; apodization; Fourier transformation;
and application of an exponential filter. Further details of each
process can be found in \citet{colless01}.

The highest peak in the cross-correlation function is then fitted with
a quadratic equation to obtain its position and height. The ratio of the peak
height to the noise level in the cross-correlation function is
subsequently computed and used to determine the quality of the
redshift estimate, via the assignment of a quality flag, Q$_{\rm a}$,
to each spectrum.  The value of Q$_{\rm a}$ ranges from 1 to 4, with 4
and 3 meaning a reliable fit, 2 a probable fit and 1 an unreliable
fit. If no fit is obtained a quality factor of 0 is assigned.  When
Q$_{\rm a}$ is 3 or 4, there is an additional requirement that at
least four or six of the eight templates, respectively, produce the
same redshift to within 600 km s$^{-1}$.  In addition, the code also
outputs which of the eight templates produced the best correlation
with each spectrum; this gives information on the object's
morphological type.

For emission spectra the code performs a Gaussian profile fit to each
line after first subtracting the continuum. Any peaks in the resultant
spectrum which are higher than 3.3 times the RMS noise are marked as
candidate emission lines. The three strongest lines in each spectrum
are then tested in pairs for line separations consistent with a known
redshift from a set of common emission lines ([O\,II], H$\beta$,
[O\,III], H$\alpha$ or [N\,II]). If a match is found then all other
emission lines that match that redshift to within 600 km s$^{-1}$ are
found, and the mean redshift is adopted as the emission redshift. If
no two emission lines are found to match a particular redshift for a
pair of common lines, (or there are fewer than three lines in total)
then up to two single-line redshifts are retained for comparison with
the absorption derived redshift. A quality flag for this redshift
estimate, Q$_{\rm e}$, is assigned based on the number of features
fitted. This number can take the values 0,1,2,4 with 0 being no lines,
1 for one line, 2 for two lines and 4 for three or more.

The code then compares the redshift obtained via each method and
selects the one with the higher quality flag. In the case where
Q$_{\rm a}$ = Q$_{\rm e}$ the absorption value is selected. The best
quality flag , Q$_{\rm b}$ is then given as the greater of Q$_{\rm a}$
or Q$_{\rm e}$, with two exceptions: (1) if the difference in redshifts
derived via the two methods is less than 600 km\,s$^{-1}$ then Q$_{\rm
b}$ is the greatest of Q$_{\rm a}$, Q$_{\rm e}$ or 3 and (2) if Q$_{\rm
a}$ and Q$_{\rm e}$ are both less than 2 and the difference in
redshifts derived via the two methods is greater than 600 km\,s$^{-1}$
then Q$_{\rm b}$ is set to 1.

After the automatic quality flags are assigned the spectra are then
inspected manually and interactively re-fitted if necessary. Also at
this time the observer may assign their own quality flag Q
(independent of the automatic quality flag Q$_{\rm b}$) on a scale of
1 to 5, with 5 corresponding to spectra of the highest quality.

\section{Results}

Of the 338 spectra obtained, 136 turned out to be misclassified stars.
This was later tracked down to a problem with the object discriminator
software for bright objects in the pre-release SuperCOSMOS data.
For target selection we set the lower magnitude limit to $m_{\rm
B}=15.5$, at the bright end of the magnitude distribution of cluster
members.  Unfortunately, this was too bright for the object
classification software to distinguish reliably between stars and
galaxies.

\begin{figure} \centering
   \resizebox{\hsize}{!}{\includegraphics{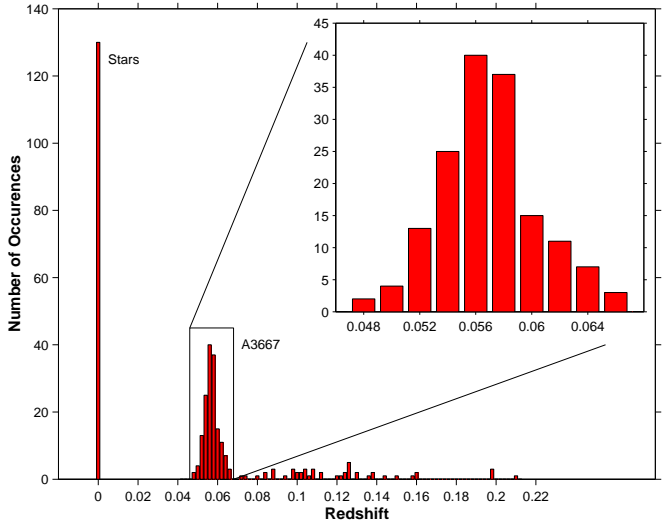}} \caption{2dF
   redshift histogram for the A3667 field; the inset plot shows an
   enlarged version of the cluster redshift distribution.}
   \label{fig:velhistwhole}
\end{figure}

Figure \ref{fig:velhistwhole} shows the distribution of redshifts
obtained from all object fibres, with bin intervals of $\Delta
z=0.002$. This figure has two main features; the misclassified stars
near zero redshift and A3667 members at around $z=0.056$.  The inset
A3667 histogram shows remarkably little velocity structure either
surrounding the main cluster group or within it. There is no evidence
of distinct in-falling groups or other velocity features around the
main cluster as found in other merging systems like the
A3125/A3128 complex \citep{Rose02}. Then again, perhaps this is not
surprising as clusters which harbour diffuse radio emission have been
shown to be statistically more isolated than clusters of similar X-ray
luminosity \citep{Schuecker99}.

\begin{figure} \centering
   \resizebox{\hsize}{!}{\includegraphics{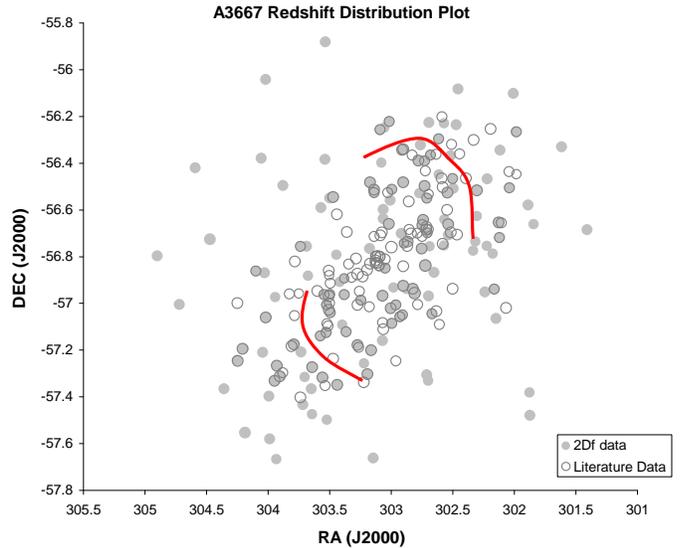}}
   \caption{Distribution on the sky of galaxies with measured
   redshifts in the region of A3667. Filled circles are the 2dF data
   and open circles are the redshifts from the literature.  Where the
   2dF and literature datasets agree in both position and redshift,
   the circle is shown filled and outlined. The arcs denote the outside
   edges of the radio relics.}  \label{fig:circles}
\end{figure}

After the stars had been removed, the remaining 202 spectra were
examined individually for quality of fit. Spectra for which the
quality flag Q$_{\rm b}$ was less than 2 were treated with caution and
18 spectra with low signal-to-noise were removed from the sample. Of
the remaining 184 spectra, 143 fell within the range $z=0.044-0.068$
and are considered to be members of A3667. This redshift-restricted
sample of 143 galaxies is published online at CDS in the format of
Table \ref{tab:z}; a sky plot of this sample, together with the 154
values from the literature, is shown in Figure \ref{fig:circles}.  The
remaining 41 galaxy spectra not related to A3667 but
obtained during these observations are also published online at CDS in
a format similar to those in Table \ref{tab:z}.

\begin{table*}
\caption{Redshifts for the first 10 galaxies$^{\mathrm{*}}$ in the
sample of 143 measured with 2dF in the cluster A3667 and falling in
the redshift range $0.044 \leq z \leq 0.068$ }
$$
\begin{array}{ccccccccc}
\noalign{\smallskip}
\noalign{\hrule}
\noalign{\smallskip}
 {\rm ID} & {\rm RA }     & {\rm Dec}    & m_{\rm B} &
z_{\rm abs} & z_{\rm em}  &    z_{\rm ave} & z_{\rm lit}& {\rm morphological} \\
&  {\rm (J2000)}&  {\rm (J2000)} & & & & & & {\rm type}\\
\noalign{\smallskip}
\noalign{\hrule}
\noalign{\smallskip}

1  & 20\:05\:37.49  & $$-$$56\:41\:06.1 & 16.89 & 0.0558 &        &         & & 2\\
2  & 20\:06\:26.98  & $$-$$56\:19\:50.9 & 18.16 & 0.0586 &        &         & & 3\\
3  & 20\:07\:22.03  & $$-$$56\:39\:44.0 & 16.41 & 0.0535 &        &         & & 2\\
4  & 20\:07\:29.09  & $$-$$57\:28\:46.2 & 18.09 & 0.0553 & 0.0553 & 0.0553  & & 5\\
5  & 20\:07\:30.08  & $$-$$57\:22\:58.6 & 16.75 & 0.0521 & 0.0518 & 0.0520  & & 3\\
6  & 20\:07\:32.50  & $$-$$56\:34\:45.9 & 17.43 & 0.0560 & 0.0557 & 0.0559  & & 4\\
7  & 20\:07\:56.06  & $$-$$56\:16\:01.3 & 16.65 & 0.0543 & 0.0545 & 0.0544  & 0.0535 & 5\\
8  & 20\:08\:01.56  & $$-$$56\:06\:07.6 & 17.60 & 0.0594 & 0.0594 & 0.0594  & & 3\\
9  & 20\:08\:09.17  & $$-$$56\:30\:21.7 & 16.80 & 0.0523 &        &         & 0.0524 & 2\\
10 & 20\:08\:27.61  & $$-$$56\:20\:42.5 & 17.79 & 0.0572 & 0.0569 & 0.05705 & & 3\\
\noalign{\smallskip}
\noalign{\hrule}
\end{array}
$$
\begin{list}{}{}
\item[$^{\mathrm{*}}$]
The full data table contains 143 objects and is available in electronic
form at CDS via anonymous ftp to cdsarc.u-strasbg.fr (130.79.128.5)
or via http://cdsweb.u-strasbg.fr/cgi-bin/qcat?J/A+A/(vol)/(page). Column 1
gives the source ID number; columns 2 and 3 are the Right Accession and 
Declination in J2000 coordinates; column 4 is the blue magnitude; columns 6 
\& 7 are the redshifts as determined by absorption and emission features
respectively and column 8 is the average redshift for cases where a redshift
could be determined by both emission and absoprtion features; column 9 is
the previously published redshift and column 10 is the morphological type 
determined by the best correlation to the spectral templates used by
2dfdr, numbers correspond to those listed in \cite{colless01}.
\end{list}
\label{tab:z}
\end{table*}

\begin{figure} \centering
   \resizebox{\hsize}{!}{\includegraphics{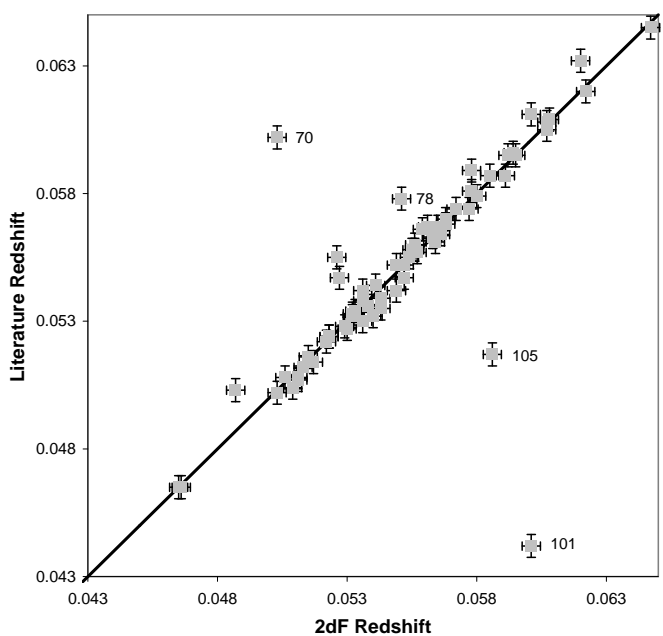}}
   \caption{Redshift comparison for A3667. The literature redshift is
   plotted against the 2dF redshift for the 67 sources in common.
   Clustering about the slope of unity suggests there is no overall
   systematic difference between the two datasets. The four labelled
   outliers are discussed in the text; the number corresponds to its
   identification number in the online version of Table \ref{tab:z}.}
   \label{fig:zerror}
\end{figure}

There are 67 galaxies in common between the 2dF and literature
samples.  As can be seen from Figure \ref{fig:zerror}, there is
generally good redshift agreement between the two datasets. 
Analysis of the differences in redshift gives a median difference of
zero and a standard deviation of $\Delta z=0.00036$ (102
km\,s$^{-1}$), well below the combined redshift uncertainty ($\sim$
200 km\,s$^{-1}$). This suggests that it is acceptable to combine the
two datasets.

In merging the two datasets we took note of the 2dF quality
parameters.  The {\it blunder\/} rate for spectra measured in the
2dFGRS for which \verb+2dfdr+ has assigned Q$_{\rm b}$ $>$ 2, is only
1.6\% and this depends strongly on magnitude, with almost all blunders
occurring for $m_{\rm B}\geq 18$ \citep{colless01}. Thus, for a
comparatively bright sample such as this, one can have confidence in
the reliability of the 2dF redshifts.  When combining the two datasets
for subsequent analysis the 2dF redshifts were used in preference to
the literature values in all cases where Q$_{\rm b}$ was greater than
two. Of the nine 2dF spectra with Q$_{\rm b} \leq 2$, seven agreed well
with the previously published redshifts and were preferred for the
combined dataset.  The remaining two spectra showed large
discrepancies of $\Delta z =0.0069$ and 0.0027 (corresponding to
$\sim$ 2000 and 800 km\,s$^{-1}$ respectively).  These points can be
seen on Figure \ref{fig:zerror} as outliers numbered 105 and 78 (which
correspond to their identification numbers in Table \ref{tab:z}).
Both objects were quite bright with magnitudes of 16.90 and 17.39, and
quoted errors of 61 and 200 km\,s$^{-1}$ respectively \citep{Sodre92}; for
consistency, the 2dF values were retained for analysing the redshift
distribution.  The remaining two labelled outliers, 70 and 101, were
assigned the highest quality flags by \verb+2dfdr+ and were thus
assumed to be correct.  This gave a combined dataset of 231 redshifts
in the range 0.044 to 0.068, which were then used in the subsequent
analysis.

\begin{figure} \centering
   \resizebox{\hsize}{!}{\includegraphics{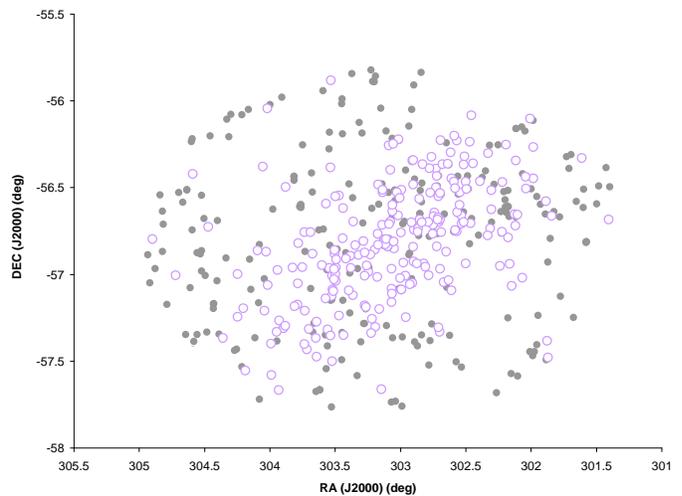}}
   \caption{Coverage of galaxies for which spectroscopic redshifts have been
determined. Filled gray circles are all the galaxies down to
  b$_j$ $\leq$ 18.3 in a one degree radius about A3667 as determined from SuperCOSMOS 
plate scans while open circles represent those galaxies for which redshifts have been
obtained. The average completeness over the field is 70\%.}  \label{fig:coverage}
\end{figure}

It should be noted that these observations combined with those previously
published mean that A3667 has been spectroscopically observed at a completeness
level of 70\% down to our limiting magnitude of b$_j$ =18.3. Figure 
\ref{fig:coverage} shows all galaxies in a field of one degree radius centred on the peak in
X-ray emission down to b$_j$ =18.3 as gray points overlaid with all galaxies for which 
spectroscopic redshifts are currently available.

\subsection{Redshift Distribution}
\label{reddist}

 \begin{figure} \centering
   \resizebox{\hsize}{!}{\includegraphics{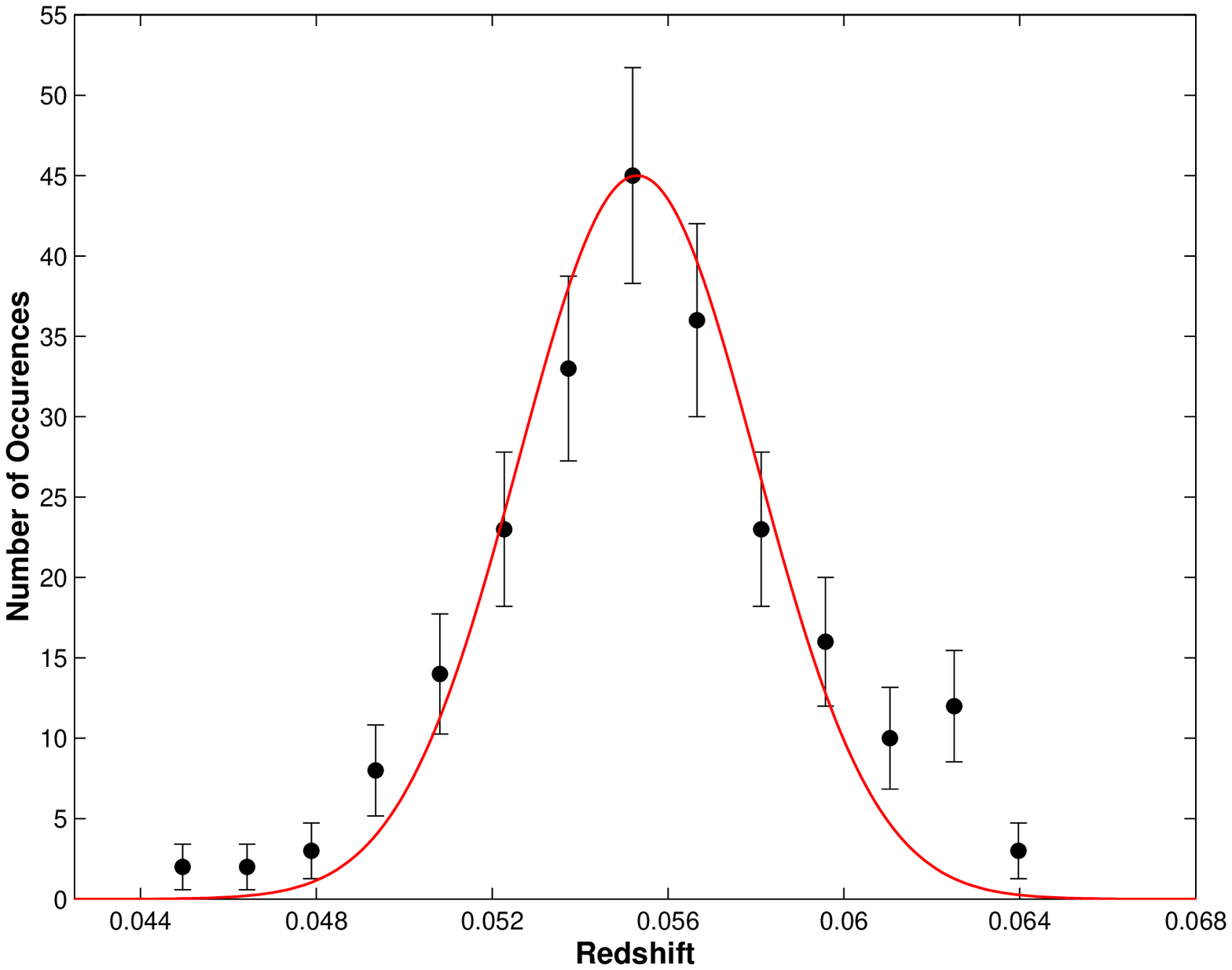}}
   \caption{Redshift distribution for A3667 with a Gaussian fitted to
   the central peak.  The Gaussian is centred at the mean redshift of $z=0.0555$ and there
   is a weak excess at around 0.0625. The velocity dispersion is 1102 km
   s$^{-1}$.}  \label{fig:velcurv}
\end{figure}

\begin{figure}[htbp]
\centering 
\resizebox{\hsize}{!}{\includegraphics{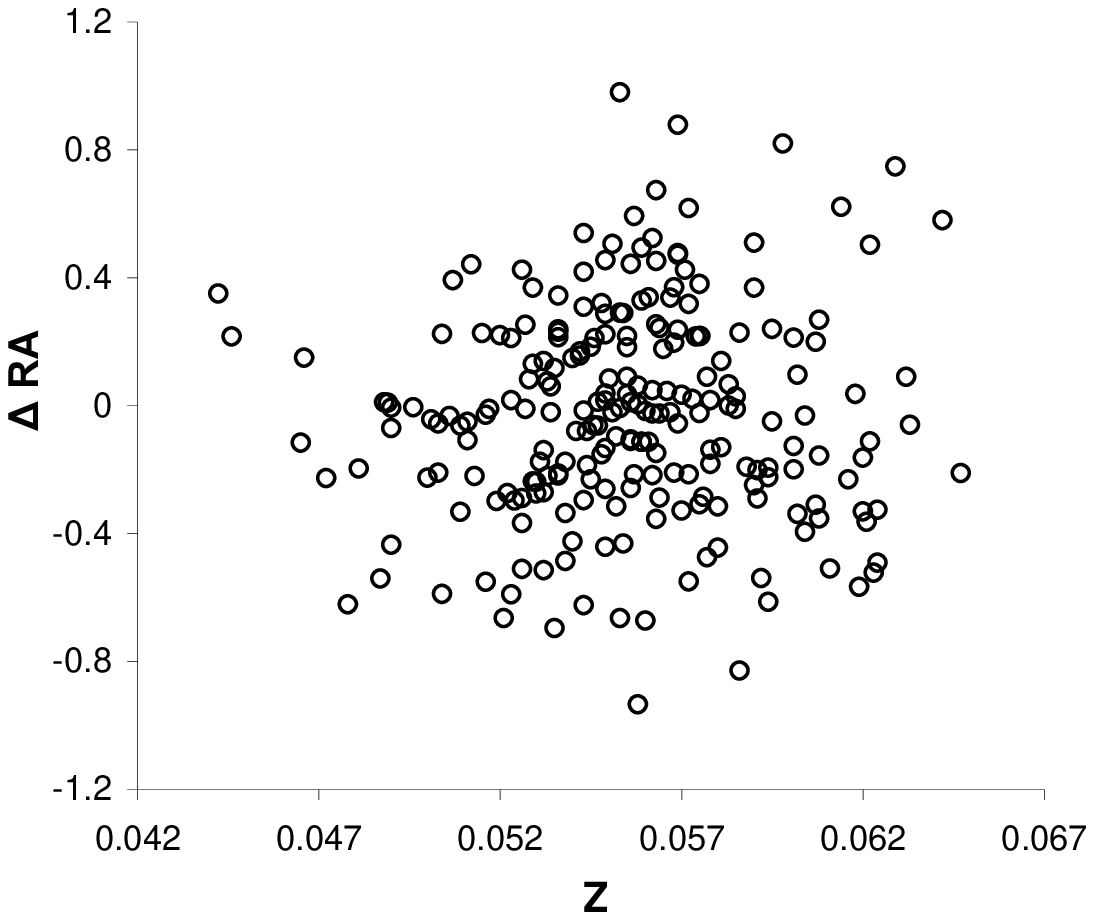}}
\resizebox{\hsize}{!}{\includegraphics{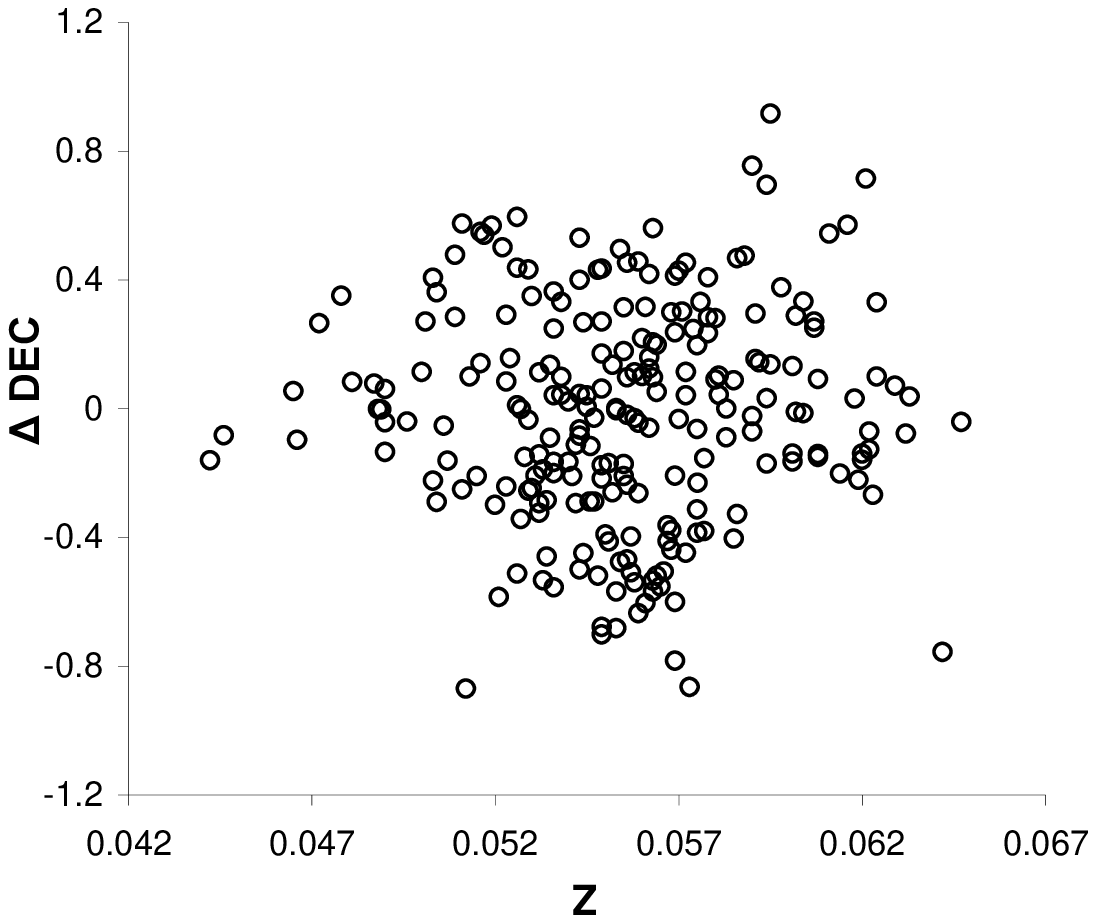}} 
\caption{Redshift distribution for the combined dataset of 231
galaxies as a function of RA and Dec, measured in degrees from the
cluster centre.}
\label{fig:assoc}
\end{figure}

The redshift distribution for the combined dataset is shown in Figure
\ref{fig:velcurv}, overlaid with a Gaussian fit to the central
peak at $z=0.0555$.  
While the redshift distribution is well modelled by a Gaussian, there
is some evidence for a small subgroup of galaxies at a slightly higher
redshift; however, the galaxies belonging to this possible subgroup
are not spatially localised but are spread across the cluster.
To search for possible subgroups a pseudo-3D plot of RA, Dec and
redshift was constructed using Matlab to explore the spatial
distribution of all 231 galaxies.  While such a plot does not give a
physical 3D model of the cluster, it is a useful tool in searching
for associations.  Standard redshift versus $\Delta$RA and $\Delta$Dec
plots were also made (see Figure \ref{fig:assoc}) and tested for
associations, on the premise that if subgroups are present there will
be tight associations between the member galaxies in both plots. No
likely subgroups were found via visual inspection and the cluster 
appears extremely well mixed. This lack of velocity structure is consistent
with previous results \citep{Girardi96}.

Statistical analysis of the total dataset gives a mean redshift of 0.0555, 
and a velocity dispersion, $\sigma_{\rm v}$, of 1102 $\pm$46 km\,s$^{-1}$, 
which is slightly higher than the previously published values.

\subsection{Quantitative Tests for Substructure}
In order to produce a more quantitative analysis of the likely substructure 
present we performed 2 and 3-dimensional Lee-Fitchett tests 
\citep{Fitchett87, Fitchett88} in addition to calculating the $\Delta$ statistic
for the cluster \citep{Dressler88}.

\subsubsection{Lee-Fitchett Test}
We performed a Lee-Fitchett (LF) test \citep{Fitchett87, Fitchett88} on 
the 231 galaxies known to be members of 
A3667. The ratio obtained was 5.15 which is highly significant. However, 
as the LF test is simply a measure of spatial
variations in variance, this does not imply subclustering in the case of
A3667 which is highly aspherical. Although the Lee-Fitchett test has 
been shown to be a good indicator of 
substructure it does have limitations in that it will breakdown in the 
case of two widely spatially separated Gaussian velocity distributions. In the
case of A3667 the LF test returns a result for significant substructure 
only in the plane of the sky along the axis of elongation of the galaxy 
positions. This is to be expected given that the premise of the test is that
clusters are roughly spherical. In order to assess the veracity of the test 
we simulated several clusters of the same size as A3667 with a high degree 
of aspherity (elliptically distributed in RA-DEC) but which have a Gaussian velocity 
distribution. In all cases the simulated clusters also returned a highly significant 
substructure value for the LF test along the major axis of the ellipse 
defining the galaxy positions in RA \& DEC. Neither in the case of A3667 or 
the simulated clusters was there any significant axis of substructure
associated with the velocity data. Thus, all we can say from this test is that 
distribution of galaxies in A3667 is elongated in the plane of the sky, as 
seen readily from the RA-DEC plots.

\subsubsection{$\Delta$ Statistic}
We undertook the calculation of the $\Delta$ Statistic \citep{Dressler88}, 
which tests whether the distribution of radial velocities 
($\overline{v}$, $\sigma$) of a local ($N=15$) population of galaxies 
changes with position across the cluster. The $\Delta$ value returned for 
A3667 was 1.26.  For comparison 100,000 Monte-Carlo simulations, in which 
the positions of the galaxies were fixed in RA and Dec but randomly shuffled 
in radial velocity, returned a distribution of $\Delta$ values ranging from 
0.76 to 1.49 with a mean of 1.09 and standard deviation of 0.10. The lack of 
a clear difference from the $\Delta$ value determined for A3667 suggests 
that there is no overall correlation between position and redshift in these 
data and, therefore, no significant substructure.

\subsection{Bimodality Testing}

A series of tests were performed to pursue the issue of bimodality in
the cluster raised by P88 and S92.

\subsubsection{Velocity Gradient}

If the bimodal distribution reported by P88 is real and the two 
subgroups are at all inclined from the plane of the sky, we would
expect to observe a gradient in radial velocity across the cluster.
The Matlab cube referred to in Section \ref{reddist} was used to search
for a velocity gradient across the cluster; none was found.

\subsubsection{Kaye's Mixture Model}

To test quantitatively whether a single Gaussian is an adequate fit to
the redshift distribution shown in Figure \ref{fig:velcurv}, we used
Kaye's Mixture Model algorithm, KMM \citep{Ashman94}, which allows a
user-specified number of Gaussian profiles to be fitted to the
unbinned, raw data and provides a statistical comparison with a single
Gaussian fit.  The user provides an estimate of the mean value of each
peak and the fraction of the members in each peak.  The software, provided
by Keith Ashman, then assigns membership of each redshift to one of
the initially supplied Gaussian profiles, and returns posterior
probabilities of group membership for each profile.

The algorithm was applied for the two-profile case and Table
\ref{tab:KMM} summarises the input and output parameters. For the 231
redshifts measured for A3667, the code assigned {\it all\/} values to
be members of the main group centred about $z=0.0555$, with an overall
confidence level of 96.7\%.

\begin{table}
   \caption[KMM test results for A3667 redshift data]{KMM bimodal
fitting results for the A3667 redshift sample. }
 \begin{tabular}
 [c]{l|c|c||c|c|}\cline{2-5}%
 &\multicolumn{2}{|l||}{Input Parameters} &\multicolumn{2}{|l|}{Output Parameters}\\\cline{2-5}%
& Group1 & Group2 & Group1 & Group2 \\\hline
 \multicolumn{1}{|l|}{Mean Group z} & 0.0555 & 0.0625 & 0.0555 & ---\\
 \multicolumn{1}{|l|}{Proportion} & 0.935 & 0.065 & 1.0 & 0\\
 \multicolumn{1}{|l|}{No. of Galaxies} & 216 & 15 & 231 & 0\\\hline
 \end{tabular}
 \label{tab:KMM}
\end{table}

\subsubsection{Kurtosis and Skewness}
Tests for kurtosis and skewness were applied to the velocity data. The 
kurtosis returned a value of 0.233 which is within 1 Standard Error of 
Kurtosis (SEK) for a Gaussian distribution with the same number of 
points (SEK = 0.3223). The skewness of the velocity distribution was 
found to be 0.134 which again was within 1 Standard Error of Skewness 
(SES) for a Gaussian distribution with 231 values (SES=0.16). For skewness 
to be considered significant we would expect values approaching two. Thus 
there appears to be no significant skewness or kurtosis for the velocity
distribution as compared with a Gaussian distribution of the same number 
of data points.

\subsubsection{Isodensity Plots}

\begin{figure}[htbp]
\centering
\resizebox{\hsize}{!}{\includegraphics{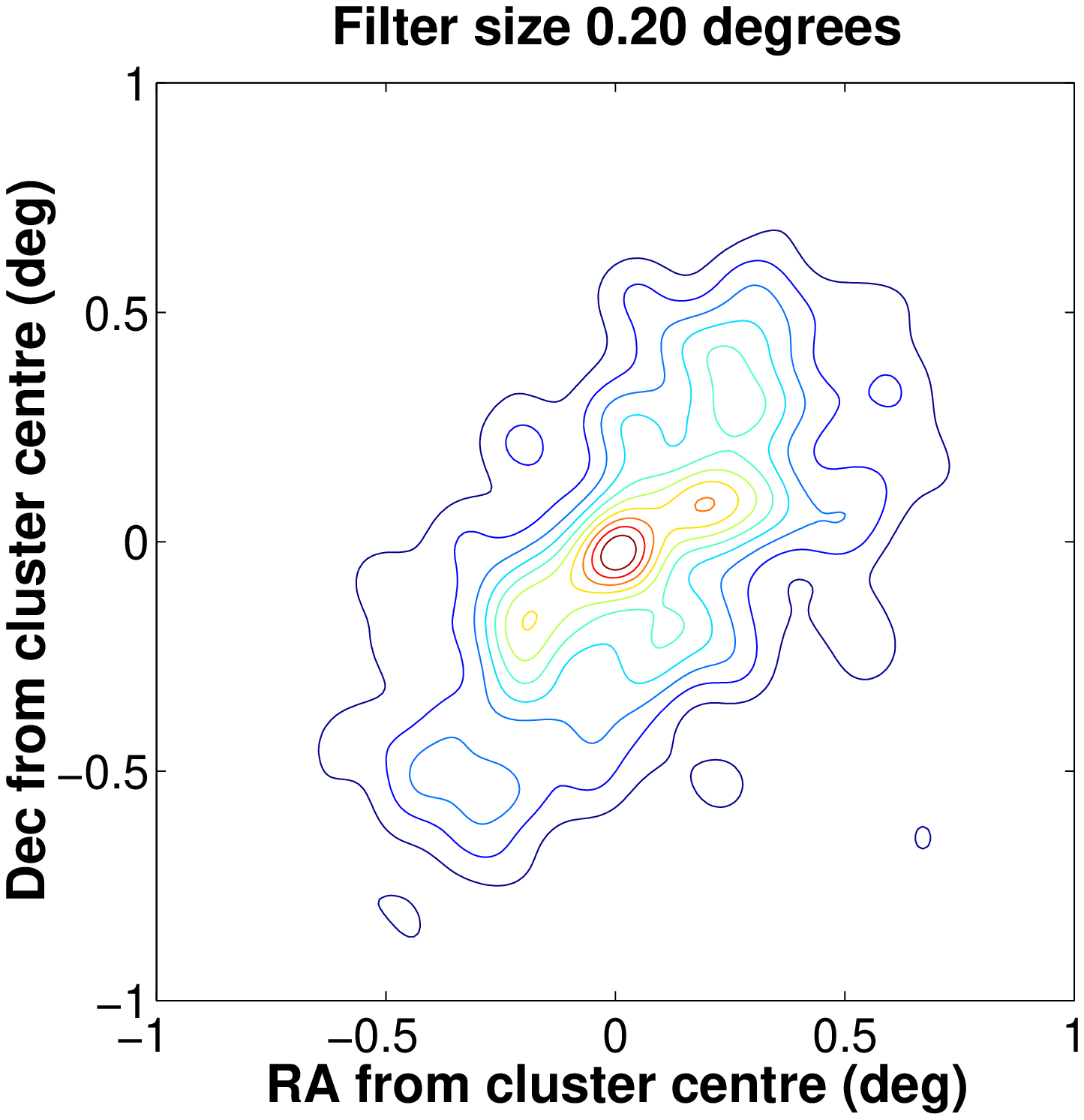}\includegraphics{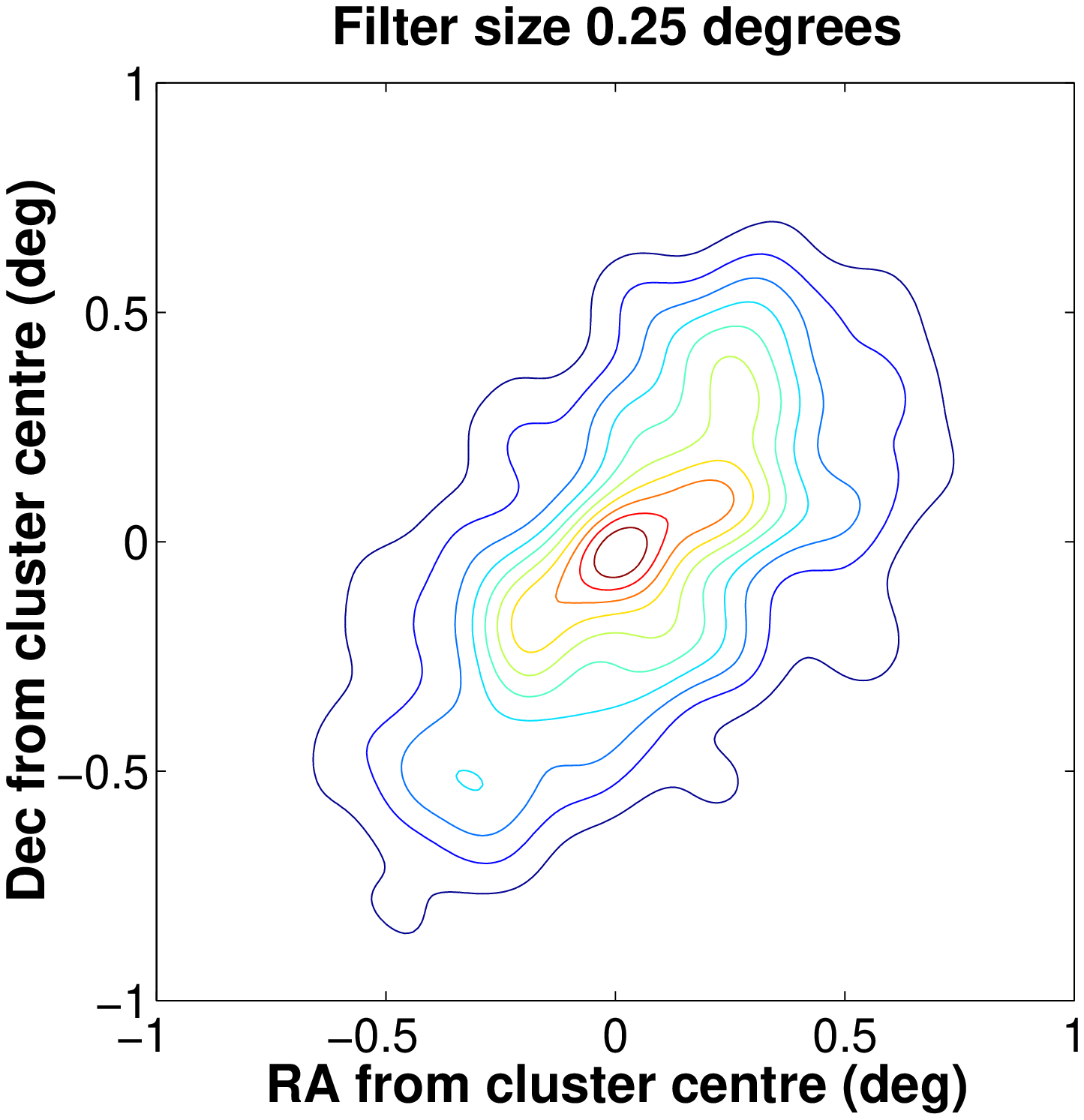}}
\centering
\resizebox{\hsize}{!}{\includegraphics{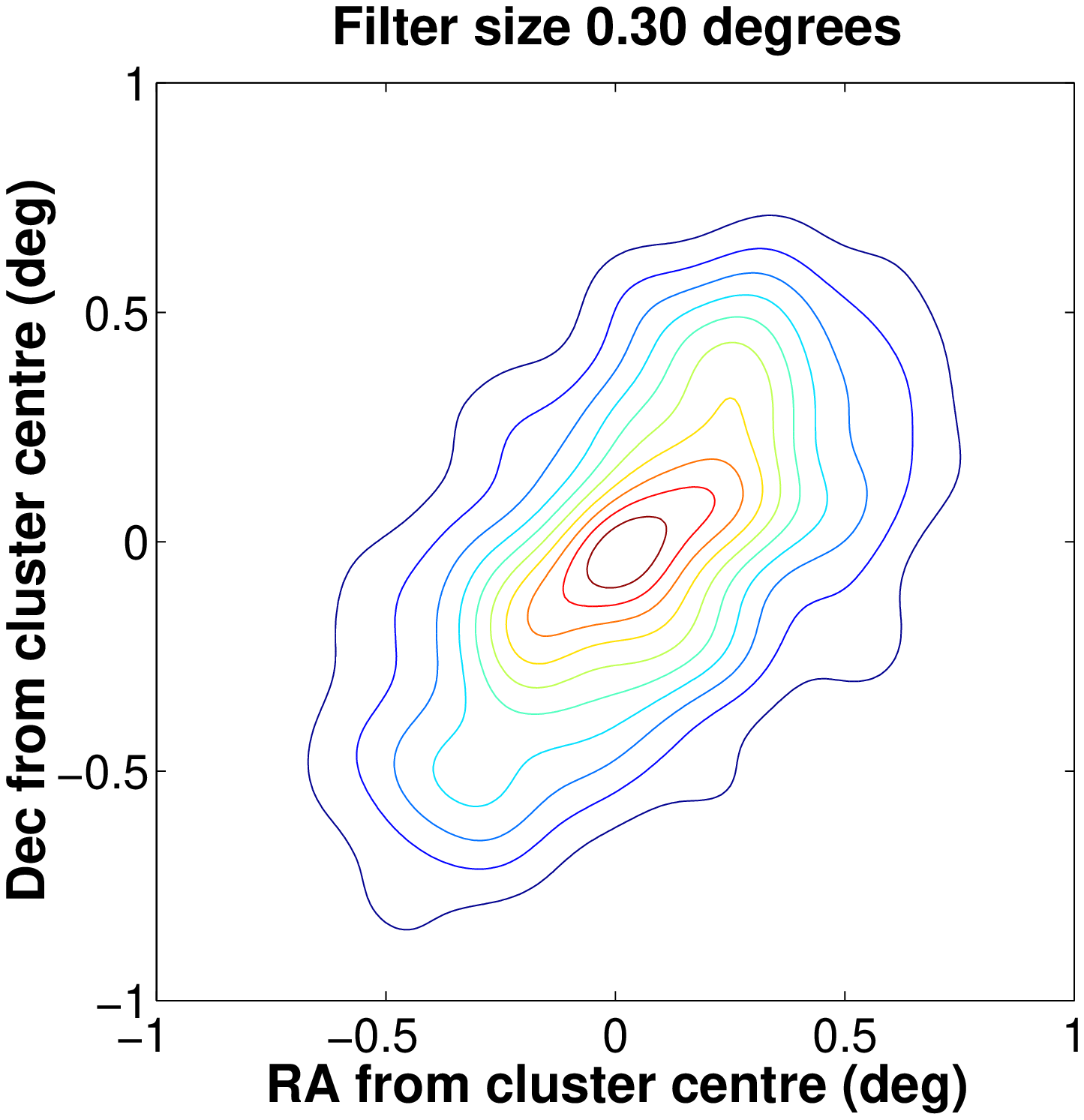}\includegraphics{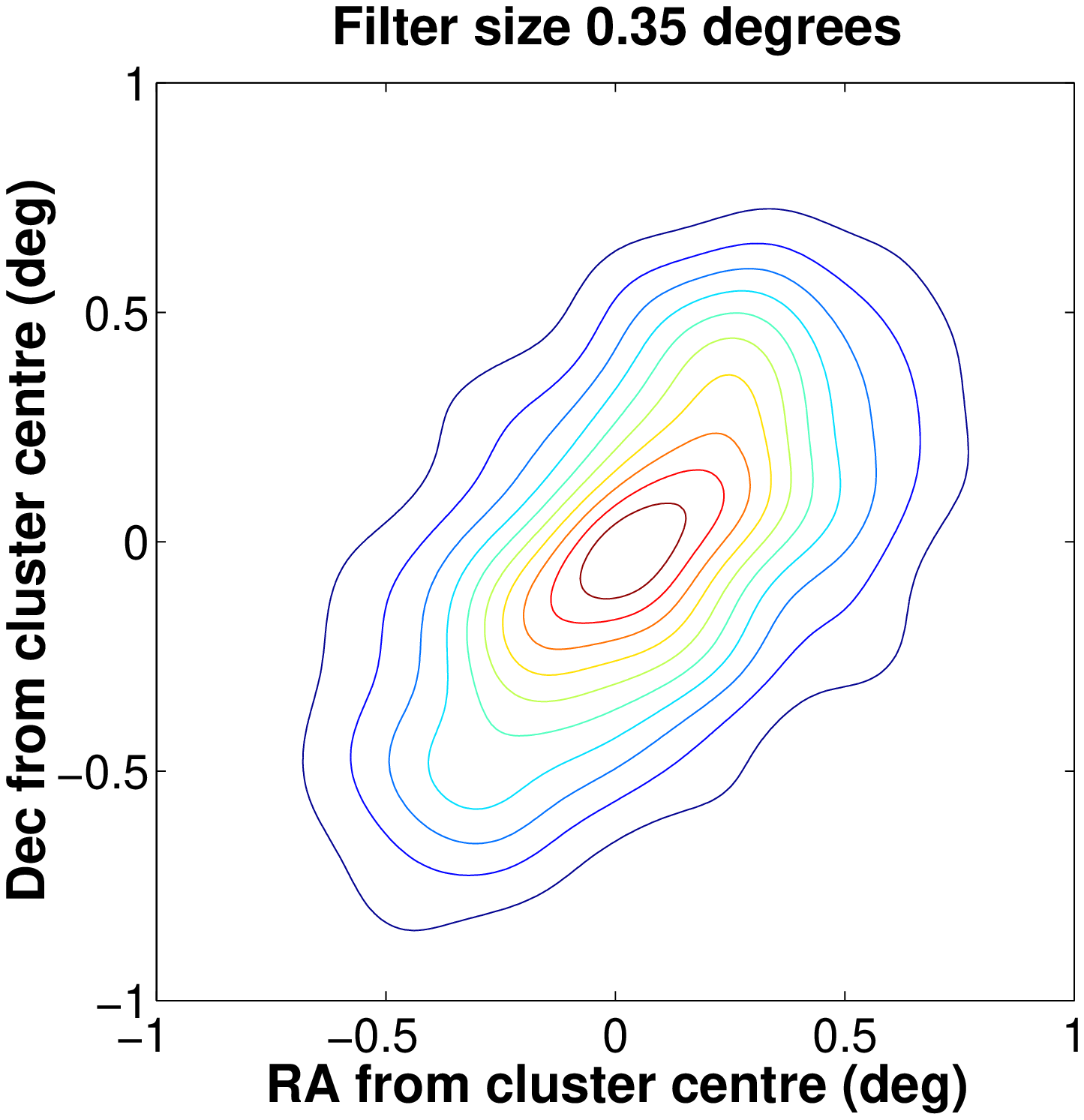}}
\caption{Galaxy isodensity plots for A3667, generated by application
of Gaussian filters of various sizes (from $\sigma = 0.2$ to
0.35$^{\circ}$) to the 231 known cluster members. Contours are shown
at $\sqrt{2}$ intervals.}
\label{fig:filterc}
\end{figure}

In order to assess the bimodality of the galaxy surface distribution,
Gaussian smoothing was applied to the spatial distribution of the 231
galaxies with redshifts between 0.044 and 0.068. Figure
\ref{fig:filterc} shows the galaxy location data smoothed with
Gaussian filters ranging in size from $\sigma = 0.20^{\circ}$ to
$0.35^{\circ}$.  The 0.20$^{\circ}$ filter image shows an excellent
overall match to the published isodensity plot of P88, except that the
redshift-restricted data do not exhibit the significant subgroup
structure seen by P88. However, there is a definite elongation of the
cluster and some evidence of a slight enhancement in galaxy density in
the NW region where P88 find a significant subgroup.  There is also a
feature of equal significance on the SE side of A3667. This second
feature is in the approximate location that \citet{Markevitch02}
predict a small subgroup based on their Chandra analysis of the X-ray
features. However, it is likely that both these features result from
the limited sample size rather than comprising distinct
subgroups. Neither feature persists if the filter size is increased
even marginally.

\subsection{Galaxy Morphologies}

As mentioned above, one of the products of the \verb+2dfdr+
cross-correlation redshift determination is the template which gives
the strongest correlation with each measured spectrum. This provides
information on each galaxy's morphological type.  Examination of the
spatial distribution of various galaxies within the cluster may
provide valuable information on the cluster's dynamical history.

\begin{figure}[htbp]
\centering
\resizebox{\hsize}{!}{\includegraphics{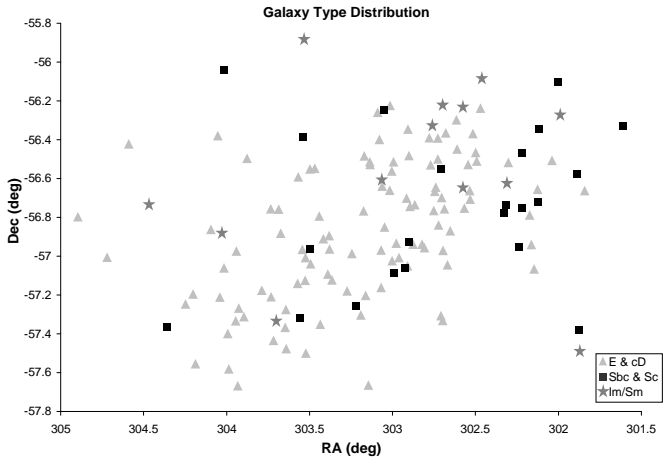}
}
   \caption{Galaxy type spatial distribution for the 143 galaxies in
the 2dF dataset.
}
   \label{fig:type}
\end{figure}

Figure \ref{fig:type} shows the distribution of galaxy types in A3667
as defined by the \verb+2dfdr+ template. The morphological types are
elliptical (E), central dominant (cD), spiral (Sc, Sbc) and
irregular/small (IM/SM).  Not surprisingly the predominant galaxy type
is elliptical. The spiral population occurs mainly in the
north-western part of the cluster and at first glance this appears
to be significant. We applied a bi-dimensional Kolmogorov-Smirnoff
test \citep{Peacock83} to test if there was a statistically
significant difference between the distribution of spiral and elliptical
galaxies in the cluster. Using positions of the 85 elliptical galaxies 
and 19 spirals with morphological types identified in the 2dF observations
we computed a P($> Z_\infty$) of 0.999 which supports the null hypothesis
that the distributions are drawn from the same population. In addition
we calculated the mean redshift and velocity dispersion for the separate
populations finding them to be almost identical with $\overline{z}_{\rm {elliptical}}$ = 0.05570
and $\overline{z}_{\rm {spiral}}$ = 0.05572 while $\sigma_{\rm {elliptical}}$=1007 $\pm$ 109 kms$^{-1}$ and 
$\sigma_{\rm {spiral}}$=1133 $\pm$ 260kms$^{-1}$.

Observations of starburst galaxies have shown a strong correlation
between star formation and tidal disruption. Such disruption may be
either the result of galaxy mergers, such as the famous case of the
Antennae galaxies (NGC4038 \& NGC4039) \citep{Whitmore99}, or from accretion or merger
shocks in the intra-cluster medium (ICM) \citep{Burns02}.  An excess of starburst 
galaxies would seem then to indicate a turbulent state, either through more
individual galaxy collisions as in the case of CL 1358+62
\citep{Bartholomew01}, or through shockwave disruption throughout the
ICM, possibly resulting in ram pressure stripping. Thus, one might
expect that an overdensity of starburst galaxies would be evidence for
a recent cluster merger. However, during the late stages of cluster
evolution we know from the Butcher-Oemler effect that cluster
constituent galaxies are more likely to be gas-stripped red
ellipticals than gas-rich blue spirals.  Without an abundance of gas
it is unlikely then that starbursts per se would be observed. A more
likely indicator of recent merging might be an excess of the so-called
`E+A' galaxies, which represent a post-starburst population
\citep{CouchSharples87} generated either through galaxy-galaxy mergers
or ram pressure stripping \citep{Belloni95}. Several studies have addressed
the prevalence of post-starburst galaxies in cluster environments 
\citep{Tran07, Hogg06, Poggianti03} as well as the field
\citep{Blake04} finding that the fraction of post-starburst galaxies is
different to the field at low redshifts (z$\leq$ 0.1). As an excess of
post-starburst galaxies may be an indicator of quenching of star formation
in galaxies as a result of the cluster environment \citep{Poggianti03} they
can be considered an important indicator or merger activity.

No E+A galaxies \citep{CouchSharples87} were found among the spectra.
Perhaps this is not surprising as only 37 examples with strong 
H$\beta$, H$\gamma$ and H$\delta$ lines
were found in a subset of over 15,000 low- to mid-redshift galaxies
observed in the 2dFGRS.  This is an incidence rate of only 0.25
percent (Deeley 2000, private communication). While at moderate
to high redshifts E+A galaxies would be more likely to be found in 
regions where galaxy-galaxy interactions are prominent, i.e., the 
central cluster region \citep{Tran03} at redshifts less than one
they tend to be found in the field \citep{Blake04} and only a few
have to date been seen in low redshift clusters \citep{Galaz00, CandR97} like 
A3667. Due to the spacing limitations imposed by the
physical size of the 2dF fibre buttons the central region of A3667
is undersampled by 2dF, as can be seen in Figure \ref{fig:circles} meaning
that if there were E+A galaxies in the cluster we would potentially miss 
them. 

\section{Discussion \& Conclusions}
Cluster mergers are rarely established based solely on evidence from one
wavelength regime, generally requiring a multiwavelength approach to 
reveal the true dynamical nature of the interactions. 

The presence of diffuse radio emission in clusters of galaxies points
to the existence of cluster-wide magnetic fields of the order of 0.1 --1
$\mu$G and a population of relativistic electrons with Lorentz factor, 
$\gamma \gg $ 1000 and energy densities of 10$^{-14}-10^{-13}$ 
erg cm$^{-3}$ \citep{Feretti04}. Diffuse radio emission is observed most
commonly in clusters which are believed to have undergone a recent merger
event and this leads naturally to the conclusion that the phenomenon is
due to turbulence and shocks in the intra-cluster medium energizing the
electron population \citep{Harris80, Tribble93, Giovannini02}. 
Simulations of the turbulence 
and shocks in the cluster environment are able to predict the 
characteristics of observed diffuse radio emission \citep{Roettiger99, 
Miniati01, Pfrommer07} and studies of the number counts
suggest that the most likely lifetimes of diffuse radio emission is of
the order of 1 Gyr \citep{Kuo04}. Additionally, theoretical models of 
merging systems suggest that X-ray substructure will be erased after 
the sound crossing time and thus, clusters with disturbed X-ray 
isophotes are likely to be dynamically young and in on-going merger 
events. Finally, the strongest evidence for
clusters being in recent dynamical interactions is the detection of
temperature changes across the source \citep{Govoni04, Pratt07}. 

While the number of clusters containing diffuse emission is still 
comparatively low \citep{Feretti04}, there are now a number of well
studied systems with multiwavelength data 
including A2256 \citep{Clarke06}, A754 \citep{Henry04}, IC1262 \citep{Hudson04}
as well as the numerous examples presented by \citet{Govoni04}. 
To date most analysis tends to concentrate 
on the radio and X-ray properties of the clusters. However, often
this is insufficient to determine the merger geometry. For example, the case of 
A3921 \citep{Ferrari06} required data from optical,
radio and X-ray analysis to determine the dynamical state of the cluster. In 
this instance optical observations alone were inconclusive but taken together
with the other wavelength data the merger axis could be established.
Similarly, high resolution X-ray from XMM-Newton analyzed in conjunction 
with existing multiwavelength data has proved vital to unraveling the cluster dynamics in several cases such as A85 \citep{Durret05} and A3921 \citep{Belsole05}.

A3667 has a mean redshift of 0.0555, a high velocity dispersion (1102 $\pm$ 46
km\,s$^{-1}$) and an elongated optical axis.  It is typical of rich
clusters in that it is dominated by elliptical galaxies. There is marginal evidence
for a grouping of spirals in the north-western part of the
cluster. The velocity distribution is well modelled by a single
Gaussian peaking at $z=0.0555$, with a weak excess around a redshift
of 0.0625. There seems to be little or no support for distinct
in-falling groups either from the velocity information or the spatial
distribution of galaxies with measured redshifts. Moreover, there is
no evidence for a velocity gradient across the cluster suggested by
the previously reported bimodal optical distribution. However, A3667
is a rich cluster and to date only 231 redshift have been obtained for
cluster members.  A larger sample is needed to categorically rule out
the presence of substructure.

Several authors have claimed that A3667 is observed in a post-merger
state \citep{Rottgering97, KHB, Roettiger99, Markevitch99, Vikhlinin01}. 
This may be consistent with the high velocity dispersion and
mixing in the cluster. If the observed diffuse radio emission was
generated via a cluster-cluster merger it is expected that the
collisional velocities would be a few thousand km\,s$^{-1}$. If such a
merger occurred, the current data suggest that the merger axis must be
very close to the plane of the sky. This would partly account for the
lack of observable subgroup peaks in the velocity data.

Full understanding of the the dynamical situation in A3667 will require
multi-wavelength analysis. We have obtained detailed new radio data for
the entire cluster at 1.4, \& 2.4 GHz, we will use this in conjunction
with previously published radio (843 MHz) and X-ray data to further 
investigate this cluster. Findings will be published in a future paper.

\section*{Acknowledgements}

We gratefully acknowledge the 2dfGRS team for use of the \verb+2dfdr+
pipeline code, and especially Dr Terry Bridges and Dr Russell Cannon
for help with the observations and reduction.  We thank Dr Keith
Ashman for use of his mixture model analysis code, Christopher Hollitt
for help with the Matlab code, and Dr Jim Caswell and Dr
Michiel van Haarlem for constructive comments on the manuscript. 
We thank the referee for the suggestion of several statistical tests and
finally we thank the editor John Beckman for his extreme patience in dealing
with the lengthy delays associated with this manuscript.

\bibliographystyle{aa}
\bibliography{mjh_2df1_ref}

\end{document}